\documentclass[twocolumn,prd,noshowpacs,nofootinbib,amsmath,amssymb,superscriptaddress,preprintnumbers]{revtex4}
\usepackage{graphicx,epsfig,psfrag,bm,amssymb}
\usepackage{dcolumn}
\usepackage{bm}
\usepackage{color}
\usepackage{mathrsfs,amsmath, amsfonts,hepunits, color}
\usepackage{tikz}
\usepackage{slashed}
\usepackage{cancel}
\usepackage{mciteplus}
\usepackage{verbatim}
\usetikzlibrary{arrows,shapes}
\usetikzlibrary{trees}
\usetikzlibrary{matrix,arrows}

\usepackage[utf8]{inputenc}											
\usepackage{slashed}

\newcommand{\be}{\begin{eqnarray}}
\newcommand{\ee}{\end{eqnarray}}
\def\lsim{\mathrel{\rlap{\lower4pt\hbox{\hskip 0.5 pt$\sim$}}
    \raise1pt\hbox{$<$}}}                
\def\gsim{\mathrel{\rlap{\lower4pt\hbox{\hskip1pt$\sim$}}
    \raise1pt\hbox{$>$}}}

\def\lsim{\mathrel{\rlap{\lower4pt\hbox{\hskip1pt$\sim$}}
    \raise1pt\hbox{$<$}}}
\def\gsim{\mathrel{\rlap{\lower4pt\hbox{\hskip1pt$\sim$}}
    \raise1pt\hbox{$>$}}}

\newcommand{\mev}{{\rm MeV}}
\newcommand{\gev}{{\rm GeV}}

\newcommand\brabar{\raisebox{-4.0pt}{\scalebox{.2}{\textbf{(}}}\raisebox{-4.0pt}{{\_}}\raisebox{-4.0pt}{\scalebox{.2}{\textbf{)}}}}

\begin{document}

\title{Illuminating New Electroweak States at Hadron Colliders}
\author{Ahmed Ismail}
\affiliation{University of Illinois, 845 W Taylor Street, Chicago, IL 60607, USA}
\affiliation{Argonne National Laboratory, 9700 S Cass Avenue, Argonne, IL 60439, USA}
\author{Eder Izaguirre}
\affiliation{Perimeter Institute for Theoretical Physics, 31 Caroline St. N, Waterloo, ON, N2L 2Y5, Canada}
\author{Brian Shuve}
\affiliation{SLAC National Accelerator Laboratory, 2575 Sand Hill Road, Menlo Park, CA 94025, USA}

\preprint{SLAC-PUB-16519}

\begin{abstract}
In this paper, we propose a novel powerful strategy to perform searches for new electroweak states. Uncolored electroweak states appear in generic extensions of the Standard Model (SM) and yet are challenging to discover at hadron colliders. This problem is particularly acute when the lightest state in the electroweak multiplet is neutral and all multiplet components are approximately degenerate. In this scenario, production of the charged fields of the multiplet is followed by decay into nearly invisible states; if this decay occurs promptly, the only way to infer the presence of the reaction is through its missing energy signature. Our proposal relies on emission of photon radiation from the new charged states as a means of discriminating the signal from SM backgrounds. We demonstrate its broad applicability by studying two examples: a pure Higgsino doublet and an electroweak quintuplet field.
\end{abstract}

\maketitle

\section{Introduction}
\label{sec:intro}
One of the primary goals of Run 2 of the Large Hadron Collider (LHC) is to study the extent of the naturalness of the electroweak scale. The stability of the electroweak scale under radiative corrections typically requires the presence of new, massive states that 
communicate with the Higgs sector of the Standard Model (SM). When new colored particles are within kinematic reach of the LHC, the physics of naturalness is most readily probed via their strong production and subsequent striking decays:~for instance, missing energy signatures can be obtained from the decays of colored particles while the dominant backgrounds are electroweak. If the accessible new states carry only electroweak charges, however, the relatively low signal rates are swamped by SM weak backgrounds, making discovery of the new particles challenging at the LHC. Dedicated search strategies are needed to ensure that these signatures of new physics are not missed.

New electroweak multiplets are ubiquitous in theories beyond the SM. Because many vector-like electroweak states have neutral components in the multiplet, they are simple and natural candidates for dark matter (DM) \cite{Jungman:1995df,Cirelli:2005uq,Cheung:2012qy}. New electroweak states are also predicted by solutions to the hierarchy problem such as supersymmetry (SUSY) \cite{Dimopoulos:1981zb}; the most natural implementations of SUSY typically must have light Higgsinos (see, for example, Refs.~\cite{Brust:2011tb,Papucci:2011wy,Hall:2011aa}), and realizations of anomaly-mediated SUSY breaking \cite{Randall:1998uk,Giudice:1998xp} and split SUSY \cite{ArkaniHamed:2004fb} also feature new electroweak states as the lightest new particles.

If there exists a new multiplet of $\mathrm{SU}(2)_{\rm L}$ that is parametrically lighter than other new states,  the different components of the multiplet are nearly degenerate due to electroweak symmetry and are split only due to radiative corrections from weak gauge bosons \cite{Thomas:1998wy} and direct couplings to the Higgs field. Typical radiative mass splittings between components of the multiplet are less then 1 GeV, with the neutral component of the multiplet ($\chi^0$) generally being the lightest; for a single new fermion multiplet, tree-level splittings from electroweak symmetry breaking arise only from higher-dimensional operators and are suppressed by the cut-off scale. When the heavier, charged components of the multiplet ($\chi^\pm$) are produced and decay at a collider, most of the energy goes into the invisible, neutral component, with only a soft charged lepton or hadron track coming out of the decay vertex. While the new particles are largely invisible to the detector, their presence can be inferred indirectly through missing momentum (see, for example, Refs.~\cite{Petriello:2008pu,Gershtein:2008bf,Cao:2009uw,Beltran:2010ww,Goodman:2010yf,Bai:2010hh,Goodman:2010ku, Fox:2011fx,Rajaraman:2011wf}) when they are produced in association with initial-state radiation (ISR). Such searches have their limitations, however; if the new particle mass scale is comparable to or only moderately heavier than the SM $Z^0$ and $W^{\pm}$ boson masses, the kinematics of the new-physics final state do not differ substantially from the SM backgrounds, and the searches are limited by the systematic uncertainties on the $Z$ and $W$ background estimates. Consequently, simple monojet searches are not expected to significantly extend sensitivity to new electroweak multiplets beyond the LEP bound  \cite{Heister:2002mn,Han:2013usa,Schwaller:2013baa,Baer:2014kya,Baer:2014cua,Low:2014cba,Barducci:2015ffa}. 

Beyond monojet searches, the prospects are somewhat better for particular ranges of mass splittings. When the charged and neutral states are extremely degenerate, $\Delta M \lesssim200$ MeV, the charged state decays on macroscopic scales, giving rise to a disappearing-track signature \cite{Feng:1999fu,Gunion:1999jr,Cirelli:2005uq,Ibe:2006de,Asai:2008sk,Buckley:2009kv}. For larger splittings, $\Delta M\gtrsim$ few GeV, the decay $\chi^\pm\rightarrow \chi^0 W^{(*)}$ can produce SM states that are above detector thresholds when boosted, and soft multilepton signatures can help improve sensitivity at the LHC \cite{Gori:2013ala,Schwaller:2013baa,Han:2014kaa,Bramante:2014dza,Low:2014cba}. 

\begin{figure}[t]
\centering
\includegraphics[width=0.4 \textwidth ]{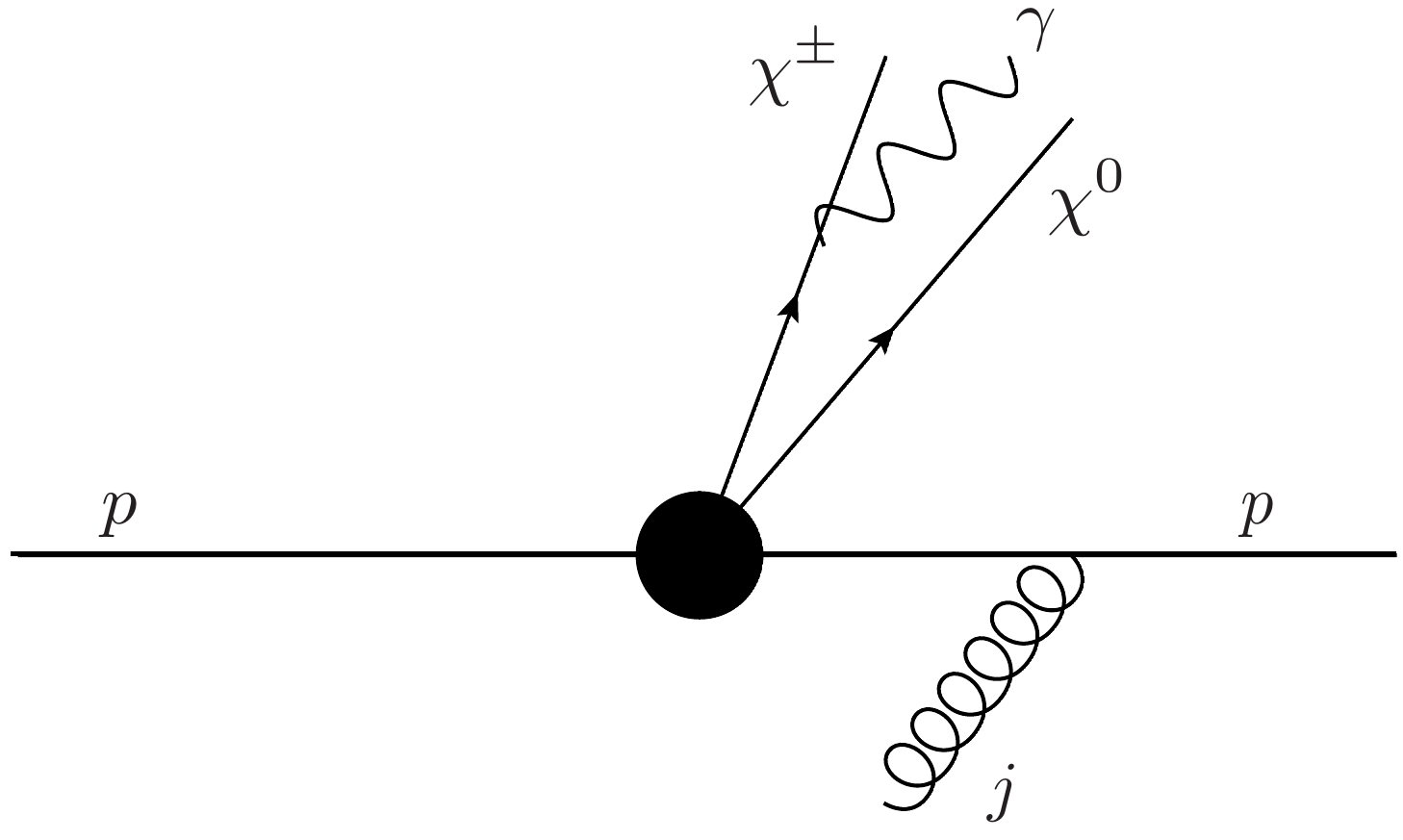}
\caption{The final state advocated in this article. Photons radiated by the charged state of the multiplet $\chi^{\pm}$ will be preferentially aligned with the missing energy in the boosted limit, in contrast with the SM $j\gamma(Z\rightarrow \bar\nu\nu)$ background.}
\label{fig:higgsinoproduction}
\end{figure}

In the intermediate mass splitting regime, $200\ \mev \lesssim \Delta M \lesssim 5\ \gev$, the SM states originating from $\chi^\pm$ decay are too soft to be identified in the detector, while the $\chi^\pm$ decay length is too short to give rise to disappearing tracks. Motivated by this challenging scenario, in this article we propose a new search strategy for discovering electroweak multiplets that is  applicable to a broad range of mass splittings. Our method relies on the fact that at least one member of any electroweak multiplet carries $\mathrm{U}(1)_\mathrm{EM}$ charge. While the $\chi^\pm$ states decay and register as missing energy from the perspective of the detector, they can still radiate soft photons prior to decay. Due to a preference for collinear emission in the highly boosted limit, such as when the $\chi^\pm$ recoils against a hard jet, final-state radiation (FSR) photons tend to be correlated with the $\chi^\pm$ momentum, and hence the missing momentum. In contrast, the corresponding SM missing momentum background from $j\gamma(Z\rightarrow \bar\nu\nu)$ does not contain photon FSR from the invisible final state particles.
Thus, we propose looking for new electroweak multiplets  by supplementing the monojet + missing energy search with a \emph{soft} photon aligned with the missing energy; such an analysis increases the signal-to-background ratio relative to the monojet search, and can provide superior sensitivity when such searches are dominated by systematic uncertainties. We illustrate the soft photon + monojet + missing energy final state in Fig.~\ref{fig:higgsinoproduction}.

We note previous work harnessing the power of electroweak FSR ({\it e.g.,} $Z$ emission) in topologies that feature neutrino final states arising from decay chains of some new coloured particles \cite{Hook:2014rka}. Additionally, previous studies have identified soft (possibly displaced) photons --- that arise from the decay of an excited state --- as a powerful handle on certain classes of DM models \cite{Weiner:2012cb,Primulando:2015lfa,Izaguirre:2015zva}, as well as the potential utility of FSR of new dark gauge bosons \cite{Cheung:2009su,Autran:2015mfa,Bai:2015nfa,Buschmann:2015awa}. Signatures involving soft photons from radiative decays in conjunction with leptons have also been considered for moderately compressed spectra \cite{Bramante:2014dza,Han:2014xoa,Bramante:2014tba}, though in contrast to these works the strategy proposed here is largely independent of mass splittings. Finally, future lepton colliders may probe nearly degenerate electroweak multiplets through their induced loop corrections to gauge boson couplings \cite{Cao:2016qgc}.

This article proceeds as follows. In Sec.~\ref{sec:models}, we introduce the two representative models that serve as benchmarks for our study. We introduce in Sec.~\ref{sec:method}  a new technique to tag these challenging final states, and in Sec.~\ref{sec:results} we show the potential for current and future hadron colliders to improve sensitivity to electroweak multiplets. We summarize our results, commenting on additional experimental background considerations, in Sec.~\ref{sec:summary}.

\section{Representative Models}
\label{sec:models}

Consider a new electroweak state with charges $(\mathbf{n},Y)$ under $\mathrm{SU}(2)_{\rm L}\times \mathrm{U}(1)_\mathrm{Y}$. Prior to electroweak symmetry breaking, all $n$ members of the multiplet are exactly degenerate. In the absence of a direct coupling to the Higgs field, the  breaking  of electroweak symmetry is communicated to the multiplet via loops of gauge bosons, which typically make the charged states heavier than the neutral ones. For a multiplet with $Y=0$, the mass splitting between the neutral state, $\chi^0$, and the states of charge 1, $\chi^\pm$, is ($M_\chi\gg M_W,\,M_Z$) \cite{Cirelli:2005uq}:
\be\label{eq:split_loop}
\Delta M \equiv M_{\chi^\pm}-M_{\chi^0} \approx \frac{\alpha_2 M_W}{2}\sin^2\frac{\theta_{\rm W}}{2}\approx 166\,\,\mathrm{MeV},
\ee
where $\alpha_2$ is the $\mathrm{SU}(2)_{\rm L}$ fine structure constant, and $\theta_{\rm W}$ is the weak mixing angle. For such a splitting, the decay $\chi^\pm\rightarrow \chi^0\pi^\pm$ is kinematically allowed, but the pion is so soft as to be essentially undetectable. The mass splittings between multiply-charged components are larger, but the loop suppression still ensures that they are $\lesssim$ GeV.

The components of $\chi$ can also have their masses split via tree-level couplings to the Higgs field. If $\chi$ is a Dirac fermion, then couplings to the Higgs arise from higher-dimensional operators such as
\be\label{eq:split_EFT}
\mathcal L \supset \frac{i}{\Lambda} (\bar\chi T^a_n \chi)( H^* T^a_2 H),
\ee
where $T^a_m$ are the generators of $\mathrm{SU}(2)_{\rm L}$ in the $m$-dimensional representation. As this gives an imaginary contribution to the real tree-level mass, the additional mass splitting scales like 
\be
\Delta M \sim \frac{\langle H\rangle^4}{\Lambda^2 M_\chi},
\ee
and is small for $\Lambda \gg 100\,\,\mathrm{GeV}$. For scalar $\chi$, there is also a direct quartic interaction between $\chi$ and the Higgs that splits the mass states. This contribution is comparable to or smaller than the loop-induced splitting in Eq.~(\ref{eq:split_loop}) for quartic couplings $\lesssim0.01$ \cite{Cirelli:2005uq}. Thus, we see that electroweak multiplets can generally give rise to nearly degenerate charged and neutral states.

If the splitting is sufficiently small, then the decay $\chi^\pm\rightarrow \chi^0\pi^\pm$  occurs on macroscopic length scales. For instance, if $\chi$ is an $\mathrm{SU}(2)_{\rm L}$ triplet with a splitting that is entirely induced by gauge boson loops, the $\chi^\pm$ has a  decay length $c\tau \approx 5.5$ cm. This is long enough to give rise to a disappearing-track signature. ATLAS and CMS have performed searches for disappearing tracks \cite{Aad:2013yna,CMS:2014gxa}, and the results are far more sensitive than searches based on missing energy alone. However, these limits on $M_\chi$ are exponentially sensitive to the mass splitting. With a small tree-level mass splitting due to higher-dimensional operators or scalar potential couplings, the disappearing-track signatures become completely insensitive to $\chi^\pm$ while the kinematics of $\chi^\pm$ production and decay are not appreciably changed. By contrast, the monojet + photon + $\slashed{E}_{\rm T}$ searches proposed in this paper are largely independent of mass splitting.

In the remainder of this section, we discuss two benchmark models that we will use in our study to demonstrate the utility of our proposed searches. For the first, we consider $\chi$ as a $Y = 1/2$ doublet, which serves as a simplified model realization of a Higgsino lightest supersymmetric particle (LSP) and is well motivated by naturalness arguments. For the second, we take $\chi$ to be in a Dirac $(\mathbf{5},0)$ representation of $\mathrm{SU}(2)_{\rm L}\times\mathrm{U}(1)_{\rm Y}$; this representation features doubly charged particles, and we quantify the gain due to the resulting enhanced photon FSR rate.

\subsection{Pure Higgsinos in Natural Supersymmetry}
\label{sec:higgsinos}
SUSY is one of the most popular solutions to the hierarchy problem. No positive signal of SUSY was found in Run 1, however, placing natural weak-scale models of SUSY in  tension with LHC results \cite{Craig:2013cxa}; early Run 2 searches have pushed limits on colored superpartners to nearly 1.8 TeV in some scenarios \cite{ATLAS-CONF-2015-067}. However, only a few superpartner masses are strongly constrained by naturalness \cite{Essig:2011qg,Brust:2011tb,Papucci:2011wy,CahillRowley:2012rv}. In particular, a natural SUSY spectrum could consist of merely Higgsinos, stops, and gluinos at mass scales in the approximate vicinity of the electroweak scale, with all other states decoupled. If the stop and the gluino are kinematically out of reach of the LHC search program, discovering Higgsinos would be a last hope for signs of naturalness.

In the limit of decoupled gauginos,
one obtains the pure Higgsino limit:~the lightest two Majorana neutralinos form a pseudo-Dirac state, while the lightest chargino has a mass comparable to the neutralinos, $M_{\tilde N^0_1} \approx M_{\tilde N^0_2} \approx M_{\tilde C_1^\pm}\approx\mu$, where $\mu$ is the Higgsino mass. In this limit, the lightest two neutralinos are essentially degenerate and the chargino is heavier by $\sim350$ MeV \cite{Cirelli:2005uq}. The chargino lifetime is too short to yield a disappearing track signature; the current search strategy is therefore to use monojet + missing energy signatures. Many studies of the monojet sensitivity to pure Higgsinos have been carried out; while various techniques and assumptions are employed, these studies are relatively consistent in their pessimism about the prospects for the LHC to substantially improve on LEP limits for Higgsinos \cite{Han:2013usa,Schwaller:2013baa,Baer:2014kya,Baer:2014cua,Low:2014cba,Barducci:2015ffa}. Given the importance of Higgsinos for natural SUSY theories, and the challenges of discovering these particles at the LHC, it is therefore a welcome possibility that a monojet + soft photon + $\slashed{E}_{\rm T}$ search could enhance the discovery prospects. If instead the gaugino SUSY breaking masses are not much larger than $\mu$, then the charged and neutral admixtures of the gauginos with the lightest Higgsino-like states can be appreciably split and give soft leptons in the decay $\tilde C_1^\pm \rightarrow W^{(*)} \tilde N_1$, which can help boost the discovery prospects for the Higgsinos \cite{Gori:2013ala,Schwaller:2013baa,Han:2014kaa,Low:2014cba}; such gains from soft-lepton tagging would always be in addition to the sensitivity of our proposed search.

\subsection{An Electroweak Quintuplet}
\label{sec:quintuplets}

Doublets and triplets under $\mathrm{SU}(2)_{\rm L}$ occur in the SM; however, larger representations are possible beyond the SM. These representations furnish particles with larger electromagnetic charges than are observed in the SM, enhancing the photon FSR signals that we highlight in this paper. In particular, a quintuplet with charge $(\mathbf{5},0)$ under $\mathrm{SU}(2)_{\rm L}\times\mathrm{U}(1)_{\rm Y}$ is motivated by minimal dark matter scenarios \cite{Cirelli:2005uq}, as the exotic electroweak quantum numbers protect the neutral component from decaying through renormalizable, tree-level operators. The phenomenology of the quintuplet model has recently been studied in Ref.~\cite{Kumericki:2012bh,Culjak:2015qja,Ostdiek:2015aga}.

We consider a Dirac quintuplet with electric charge eigenstates $\chi \equiv (\chi^{++},\chi^+,\chi^0,\chi^-,\chi^{--})$, along with their conjugate fields. In the minimal model with radiatively induced splittings between component fields, the proper decay length of $\chi^\pm$ is $\approx1.8$ cm for $M_\chi$ at or above the electroweak scale, which is long enough to occasionally leave a disappearing-track signature. The larger production cross section of quintuplet component fields therefore yields good expected sensitivity to the model from disappearing track searches, achieving an estimated mass reach of $\sim$ 600 GeV at the high-luminosity LHC \cite{Ostdiek:2015aga}; these analyses can far out-perform searches based on missing energy alone. However, in the presence of additional couplings of the quintuplet field, such as the higher-dimensional operators in Eq.~(\ref{eq:split_EFT}), the mass splitting is larger and disappearing track searches are ineffective.

Using the Higgsino doublet as a baseline, we can estimate the cross-section enhancement for a multiplet containing fields of charge $-j,\,\ldots,\,+j$. The neutral-current coupling for a state of charge $m$, $\chi_m$, in this multiplet is simply $mg_2$, while the charged-current coupling between $\chi_m$ and $\chi_{m+1}$  is $g_2\sqrt{(j-m)(j+m+1)/2}$ (where $g_2$ is the $\mathrm{SU}(2)_{\rm L}$ coupling). This is to be contrasted with the doublet;  ignoring subdominant hypercharge effects for the sake of a simple analytic estimate,  the doublet components have charged- and neutral-current couplings of $g_2/\sqrt{2}$ and $g_2/2$, respectively. The gain in rate for the quintuplet vs.~doublet for the charged- and neutral-current signal processes is found by summing over the squared couplings. In the zero hypercharge limit, the gain is  the same for both currents:
\be
&&\frac{1}{g_2^2/2}\,\left(g_2^2\sum_{m=-j}^j \frac{(j-m)(j+m+1)}{2} \right)\ \mathrm{(CC)}\nonumber\\
&=& \frac{2j}{3}\left(1+3j+2j^2\right)\\
&=&\frac{1}{2g_2^2/4}\,\left(g_2^2\sum_{m=-j}^j m^2 \right)\nonumber\ \mathrm{(NC)}.
\ee
For a quintuplet, this enhancement relative to the doublet is a factor of 20.

\begin{figure}[t]
\centering
\includegraphics[width=0.4 \textwidth ]{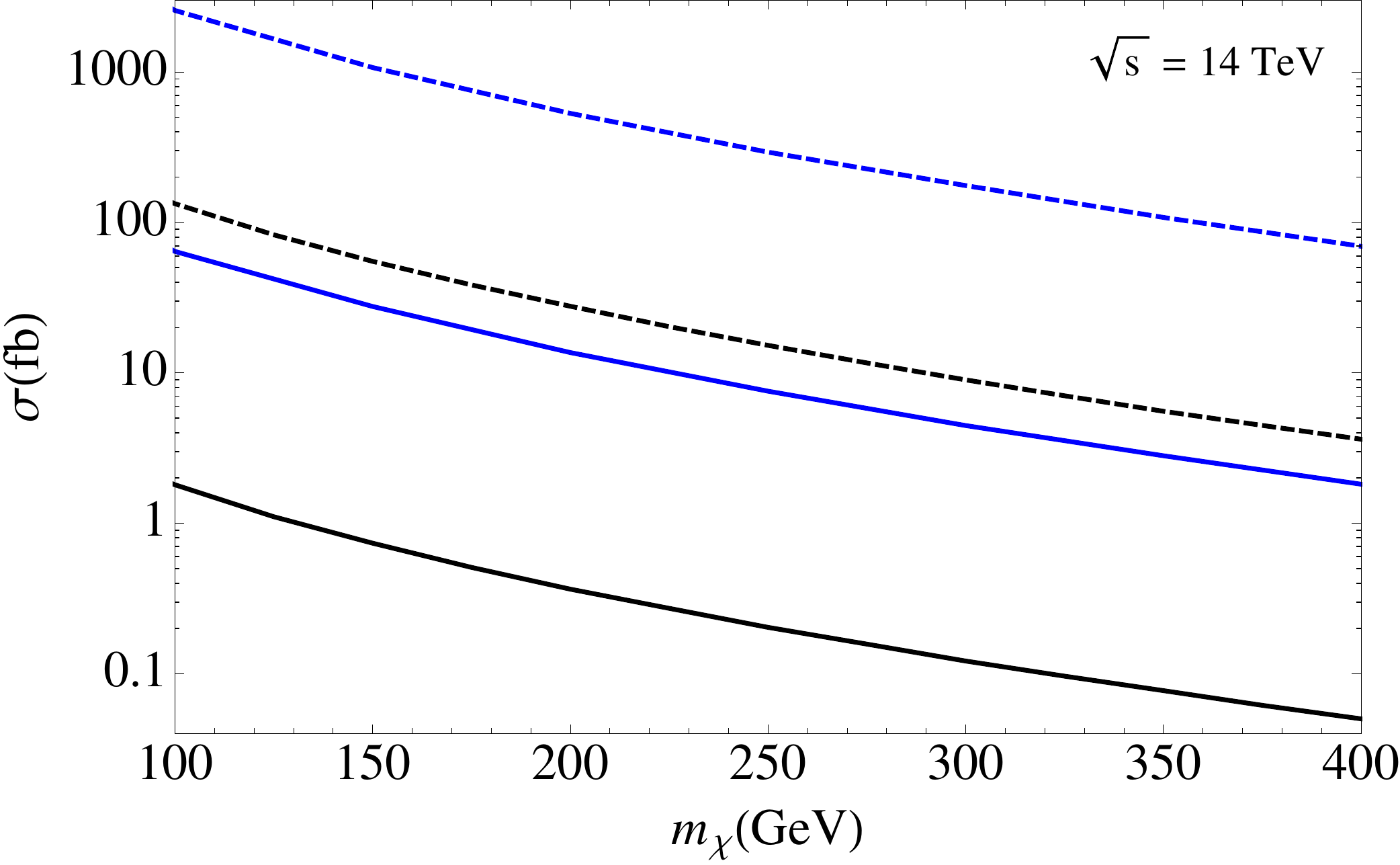}\\\vspace{0.5cm}
\includegraphics[width=0.4 \textwidth ]{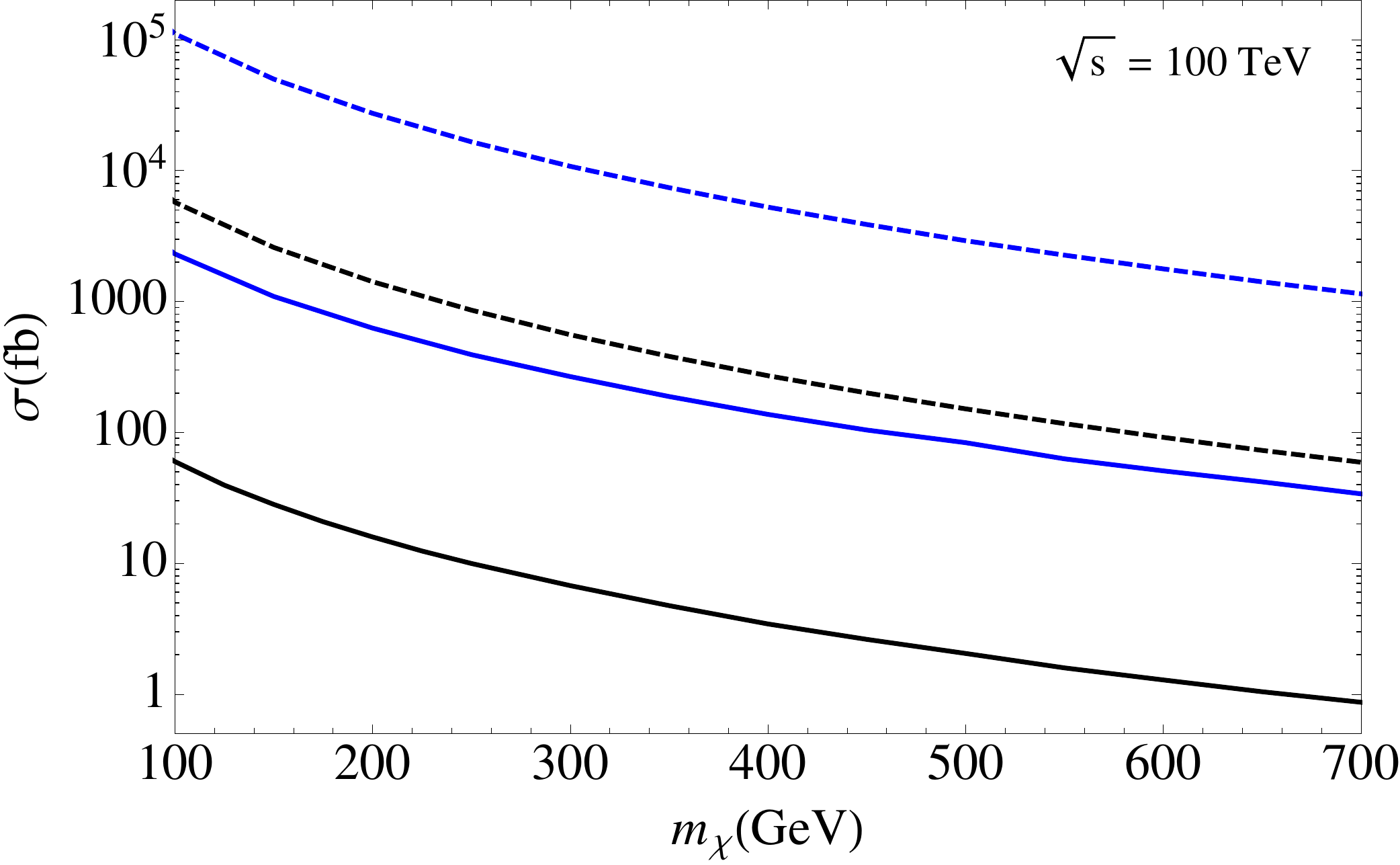}
\caption{Cross sections at 14 TeV (top) and 100 TeV (bottom) for missing momentum signatures of new electroweak states, $\chi$. We show the photon + jet + $\bar\chi\chi$ (solid) and monojet + $\bar\chi\chi$ (dashed) cross sections, matched up to one additional jet. Quintuplet production (blue) is larger than doublet production (black) because of larger $\mathrm{SU}(2)_{\rm L}$ and $\mathrm{U}(1)_{\rm EM}$ charges. We apply an NLO K-factor and impose the following kinematic cuts throughout: $p_{\rm T}(j), \slashed{E}_{\rm T} > 300$ GeV, $E_{\rm T}(\gamma) > 15$ GeV and $\Delta R(\gamma,j) > 0.4$.}
\label{fig:xsections}
\end{figure}

There is a further enhancement for quintuplet production in association with photon FSR due to the fact that some of the fields have charges $Q>1$. We expect the enhancement of photon radiation to scale like $Q^2$. For instance, we expect the rate of $\chi^{++}\bar\chi^+ \gamma$ to be enhanced by a factor of 5 relative to $\chi^+\bar\chi^0\gamma$, and for $\chi^{++}\bar\chi^{++}\gamma$ to be enhanced by a factor of 4 relative to $\chi^+\bar\chi^+\gamma$. In practice, FSR from the charged multiplet states is accompanied by ISR and photon emission from the final-state jet, and so we find the enhancements quoted above to be reduced by approximately a factor of two. Summing over final states, we find numerically that the ratio of $j+\gamma+\bar\chi\chi$ to $j+\bar\chi\chi$ is a factor of two larger in the quintuplet model than for the doublet. In Fig.~\ref{fig:xsections}, we illustrate the production cross section for $p p \rightarrow \bar\chi \chi$, including all members of the $\chi$ multiplet (charged or neutral), in association with an ISR jet with $p_{\rm T} > 300$ GeV. We also show the production cross section for the case when a soft photon with $E_{\rm T} > 15$ GeV is emitted in addition to the ISR jet, with the photon and jet separated by $\Delta R > 0.4$.

\section{A New Search for Electroweak States}
\label{sec:method}

\begin{figure}[t]
\centering
\includegraphics[width=0.4 \textwidth ]{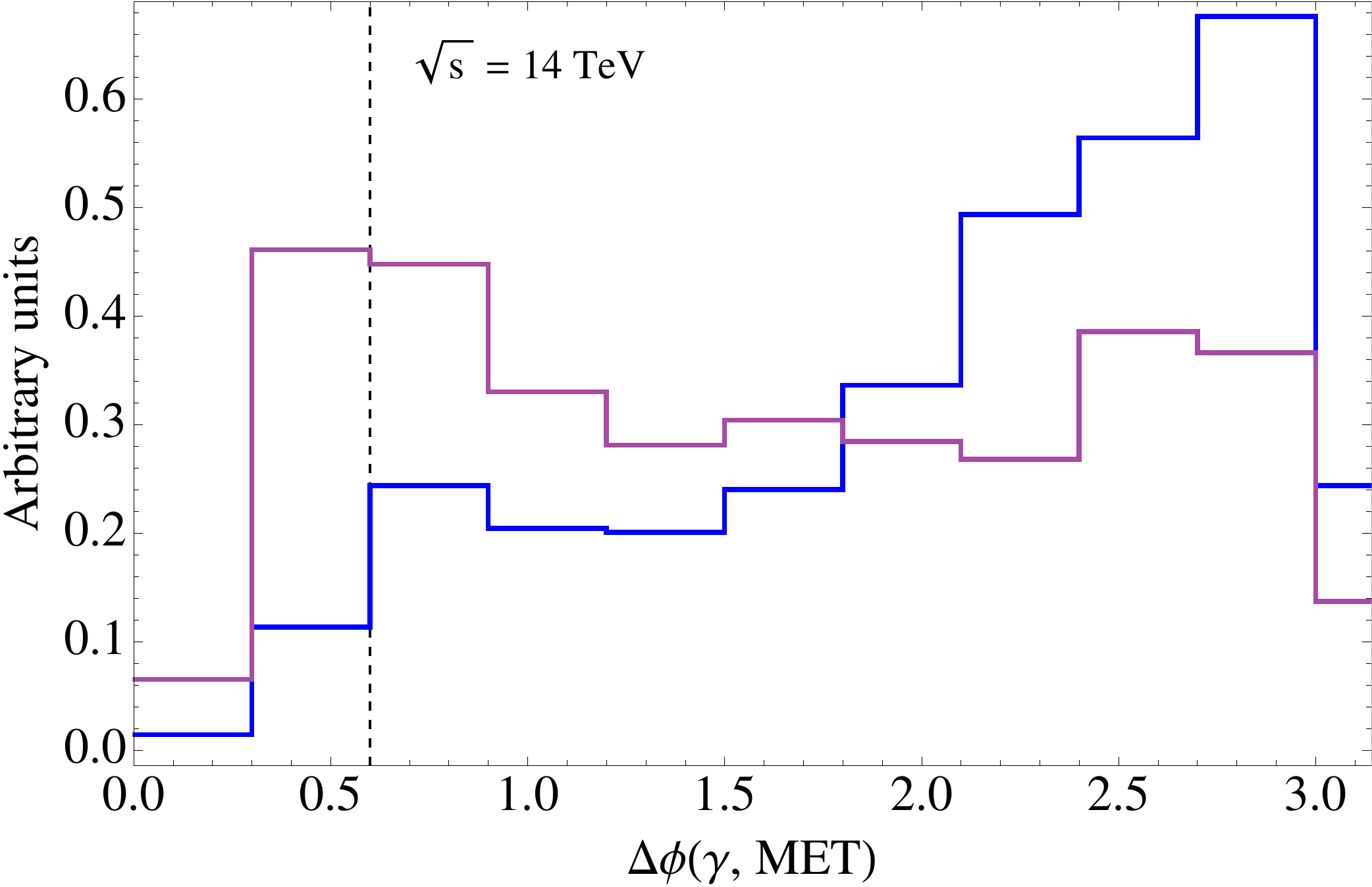}
\caption{Normalized $\Delta\phi(\gamma,\slashed{E}_{\rm T})$ distribution for an $M_\chi=125$ GeV Higgsino signal (purple) and combined $W+Z$ SM backgrounds (blue), after all selections except the $\Delta\phi < 0.6$ cut indicated in black that is optimal for this signal benchmark. Both signal and background are suppressed at $\Delta\phi=0$ due a prior  cut on $m_{\rm T}(\gamma,\slashed{E}_{\rm T})$.}
\label{fig:dphi}
\end{figure}

We now propose a new search motivated by photon radiation from the charged multiplet states;~this search is sensitive to the $\gamma + j + \slashed{E}_{\rm T}$ final state, as illustrated in Fig.~\ref{fig:higgsinoproduction}.  Because the hard jet and $\slashed{E}_{\rm T}$ are used to pass the trigger, the photon can be relatively soft, enhancing the signal rate. When the charged multiplet states are boosted, which is typically the case since the multiplet system recoils against an energetic jet,  there is a collinear enhancement of photon emission that is cut off by the mass of the charged particle. As a result, the FSR photon is often  aligned with the $\slashed{E}_{\rm T}$; this gives additional kinematic information that can be leveraged to suppress backgrounds such as $Z\rightarrow\nu\bar\nu$, which do not emit photon FSR.  We can thus expect to achieve higher signal sensitivity by imposing a requirement on $\Delta \phi \left( \gamma, \slashed{E}_T \right)$. 
As we  show in Fig.~\ref{fig:dphi} for a representative signal benchmark, the alignment between the FSR photon and missing energy can substantially improve the $S/B$ in the $\gamma + j + \slashed{E}_{\rm T}$  final state. Moreover, we find for the range of cuts considered that this gain is largely independent of that from other standard selections, such as $\slashed{E}_{\rm T}$ and $p_{\rm T}$ of the jets.
When the monojet search is dominated by systematic uncertainties, the sensitivity of the monojet + soft photon search surpasses the monojet + missing energy search alone in spite of its smaller signal rate. Additionally, the two final states are statistically independent\footnote{Monojet searches maximize $S/B$ by putting a very stringent requirement on $\slashed{E}_{\rm T}$, while the proposed search maximizes $S/B$ by requiring a soft photon and a  milder $\slashed{E}_{\rm T}$ cut; this ensures that the two signal regions have different kinematics and are largely independent samples.}, thus potentially allowing for stronger sensitivity from the combination of the two.

The main backgrounds to the $\gamma + j + \slashed{E}_{\rm T}$ search are $Z (\to \nu \bar{\nu})+ \gamma$ + jet, and $W^\pm (\to \ell^\pm + \overset{\brabar}{\nu}) + \gamma$ + jet in which the lepton is missed. In the latter case, soft photons are primarily radiated from the lepton, although we also include contributions where the photon is radiated from the $W^\pm$ prior to decay. The $W^\pm\rightarrow \ell^\pm \nu \gamma$ backgrounds are of concern because, when the lepton is missed, one obtains a soft photon aligned with the missing energy. However, we find that a cut on the transverse mass of the photon and $\slashed{E}_{\rm T}$ greatly suppresses this contribution. As in the conventional monojet searches, the top backgrounds can be controlled by requiring that the leading jet $p_{\rm T}$ is significant compared to the total missing energy \cite{Aad:2015zva}. For a sufficiently large missing energy requirement relative to the photon $E_{\rm T}$,  photon momentum mis-reconstruction is unlikely to fake $\slashed{E}_{\rm T}$. 

Less important backgrounds, such as those arising from fake photons, are still a concern in principle; however, we lack the tools to simulate these in our study. Fake photons can arise from either a lepton or a jet misidentified as a photon. For the former, the reaction $pp\rightarrow j \ell \nu$, with an electron faking a photon, could show up in the signal region. However, this final state is  removed by the same cut on the photon transverse mass, $m_{\rm T}(\gamma,\slashed{E}_{\rm T}) > m_W$, that removes real photons from $W^\pm$ decay. Regarding fake photons from jets, the final state $pp\rightarrow j j (Z\rightarrow\nu\bar\nu)$, is expected to be subdominant relative to the $pp\rightarrow j \gamma (Z\rightarrow\nu\bar\nu)$ irreducible background by far, due to small jet misidentification rates \cite{ATL-PHYS-PUB-2011-007,Bramante:2014tba}. Multijet backgrounds can be removed through a stringent missing energy requirement, and  we additionally require that any jet with $p_{\rm T} > 100\ \gev$ satisfy $\Delta\phi(j,\slashed{E}_{\rm T})>0.3$.

We perform a Monte Carlo study to assess the potential sensitivity of such a search. Parton-level events are generated at leading order with \texttt{MadGraph 5} \cite{Alwall:2014hca} and showered with \texttt{Pythia 8} \cite{Sjostrand:2007gs}; an additional jet is allowed at parton level and merged to the shower with the shower-$k_{\rm T}$ scheme \cite{Alwall:2008qv}. We use the MSSM model with all other sparticles decoupled for the Higgsino doublet model, and a  UFO model generated with the \texttt{FeynRules} package for the quintuplet  \cite{Degrande:2011ua,Alloul:2013bka,Ostdiek:2015aga}. We generate signal events $\chi\chi+j+\gamma$ for each of the models, as well as   $W/Z+j+\gamma$ backgrounds. To approximately include next-to-leading order rate effects, we apply a K-factor of 1.4 to all signal and background cross sections \cite{Denner:2011vu}. For concreteness, we study the potential reach at the high-luminosity (HL) LHC and at a future 100 TeV proton proton collider, each with an integrated luminosity of $3\,\,\mathrm{ab}^{-1}$. In our signal models, we assume only the minimal radiative mass splittings between the elements of $\chi$.

We select the following baseline signal region:
\begin{itemize}
\item At least one jet with $p_{\rm T} > 300$ GeV and $|\eta| < 2.5$;
\item  $\slashed{E}_{\rm T} > 300\ \gev$;
\item One photon with $E_{\rm T} > 15$ GeV and $|\eta| < 2.5$;
\item $\Delta \phi \left(\gamma, \slashed{E}_{\rm T} \right) < 1.4$.
\end{itemize}
The cuts on $p_{\rm T}(j)$, $\slashed{E}_{\rm T}$, $\Delta \phi \left(\gamma, \slashed{E}_{\rm T} \right)$, and $|\eta(\gamma)|$ are further optimized for each signal point over the ranges indicated in Table \ref{tab:ajmetcuts}. Events with an isolated lepton (electron or muon) with $p_{\rm T} > 7\ \gev$ and $|\eta| < 2.5$ are vetoed. Even with the lepton veto, a substantial fraction of $W^\pm\rightarrow \ell^\pm \nu \gamma$ events survive the cuts; to suppress these, we additionally require a cut on the photon transverse mass,
\be
m_{\rm T}(\gamma) \equiv \sqrt{2\slashed{E}_{\rm T}E_{\rm T}(\gamma)\left[1-\cos\Delta\phi(\slashed{E}_{\rm T},\gamma)\right]} > 80\ \gev.
\ee
 To suppress $t\bar t$ backgrounds, we require $p_{\rm T}(j_1) / \slashed{E}_{\rm T} > 0.5$. Finally, photons from pile-up contamination could be removed using timing and/or pointing information to locate the primary vertex; as these effects are detector dependent, we do not include pile-up in our analysis. We show results separately for $5\%$ and $2\%$  systematic uncertainties on the background estimate  in order to demonstrate the relative effects of statistical and systematic errors.

\begin{table}
\centering
\begin{tabular}{|c|c|c|}
\hline
& 14 TeV & 100 TeV \\
\hline \hline
$p_T(j), \slashed{E}_T$ & 300-1000 GeV & 300-4500 GeV \\
$\Delta \phi \left(\gamma, \slashed{E}_T \right)$ & 0.4-1.4 & 0.4-1.4 \\
$|\eta(\gamma)|$ & 0.5-2.5 & 0.5-2.5 \\
\hline
\end{tabular}
\caption{Ranges for optimization of cuts for $\gamma + j + \slashed{E}_T$ analysis.}
\label{tab:ajmetcuts}
\end{table}

Conventional searches for $j+\slashed{E}_{\rm T}$ have a smaller $S/B$ than our proposed analysis, but enjoy larger signal rates\footnote{Monophoton searches could also have sensitivity to new multiplets. For the  Higgsino doublet, monophoton searches are inferior to monojet searches \cite{Cirelli:2014dsa,Anandakrishnan:2014exa,Baer:2014cua}. A quintuplet is expected to radiate more photons; however, we estimate that the monophoton significance is still expected to be below the monojet significance, and we focus on the latter for our study.}. Therefore, we perform a direct comparison of the sensitivities for a monojet search with and without the soft-photon requirements. The minimum leading jet $p_{\rm T}$, missing energy, and lepton cuts are identical to those for the $j+\gamma+\slashed{E}_{\rm T}$ analysis. The jet $p_{\rm T}$ and $\slashed{E}_{\rm T}$ cuts are further scanned up to 2 (5) TeV at 14 (100) TeV to optimize the signal significance. For the $j + \slashed{E}_{\rm T}$ final state, in lieu of a cut on $p_{\rm T}(j) / \slashed{E}_{\rm T}$, we discard events with three jets with $p_T > 30\ \gev$ to suppress top backgrounds, finding that an extra jet veto offers slightly better performance.

Throughout this analysis, we assume 100\% efficiencies for photons (leptons) with $p_{\rm T} > 15\, (7)\ \gev$ and $|\eta| < 2.5$; while the currently reported average photon efficiency is $\sim60-70\%$ for soft photons \cite{ATLAS-CONF-2012-123}, this rate could be improved with a dedicated strategy for this analysis. Furthermore, these efficiencies affect both signal and background, and we have verified that the inclusion of more realistic identification efficiencies does not substantially affect our results.

\section{Results}
\label{sec:results}

In this section, we demonstrate the potential sensitivity of the search proposed in Sec.~\ref{sec:method} to the representative models discussed in Sec.~\ref{sec:models}. We compare the $\gamma + j + \slashed{E}_{\rm T}$  and monojet analyses for each model, and also show the outcome of a na\"ive combination of their sensitivities; the combination treats the samples for each analysis as independent, which is a good approximation due to their non-overlapping signal regions.

\begin{figure}[t]
\centering
\includegraphics[width=0.45 \textwidth ]{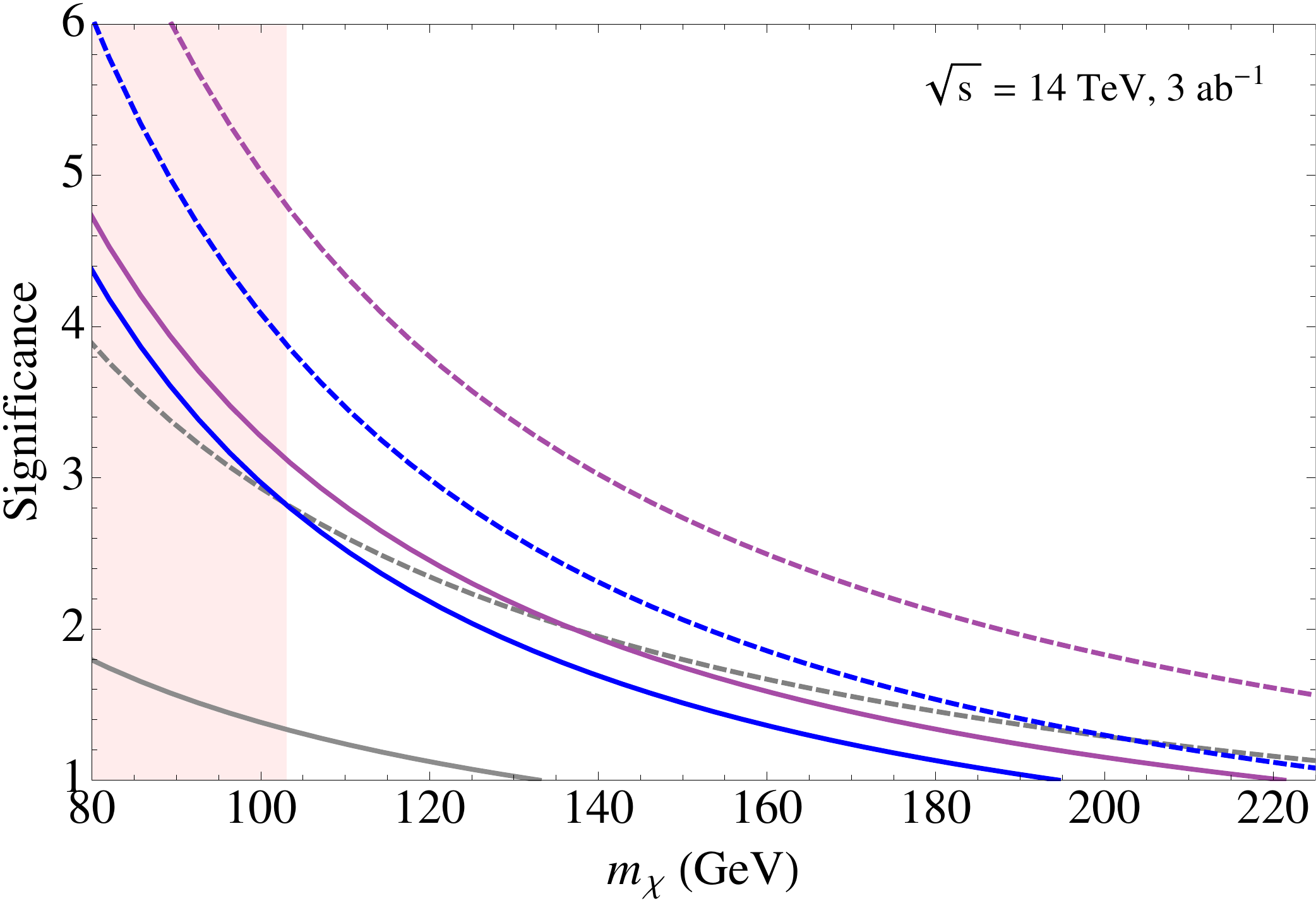}
\caption{Projected signal significance for the Higgsino doublet model with 3 ab$^{-1}$ of integrated luminosity at the HL-LHC.  Results are shown for the $\gamma+j+\slashed{E}_{\rm T}$ search (blue) assuming either 5\% (solid) or 2\% (dashed) background systematic uncertainties. The estimated $j+\slashed{E}_{\rm T}$ (gray) sensitivity is also shown for comparison, along with a na\"ive combination of $\gamma+j+\slashed{E}_{\rm T}$ and $j+\slashed{E}_{\rm T}$ sensitivities (purple). The shaded region is excluded by LEP \cite{Heister:2002mn}.}
\label{fig:moneyplot14tev}
\end{figure}

We first show the expected sensitivity of the $\gamma+j+\slashed{E}_{\rm T}$ search for Higgsino doublets at the HL-LHC in Fig.~\ref{fig:moneyplot14tev}. The search is potentially sensitive to $\sim$ 130 GeV Higgsinos at $2\sigma$ assuming 5\% systematic uncertainties on the background. The estimated reach is higher than the LEP limit of 103 GeV \cite{Heister:2002mn}, and it is a significant improvement upon the sensitivity of the monojet search alone. If systematics can be reduced to 2\%, the reach of the $\gamma+j+\slashed{E}_{\rm T}$ extends to 150 GeV. The gain in sensitivity is not as large as for the monojet search, which is more strongly systematics limited; however, the combination of the two signal regions yields a substantial improvement to model sensitivity, with a $2\sigma$ significance  for $M_\chi \lesssim190$ GeV.

\begin{figure}[t]
\centering
\includegraphics[width=0.45 \textwidth ]{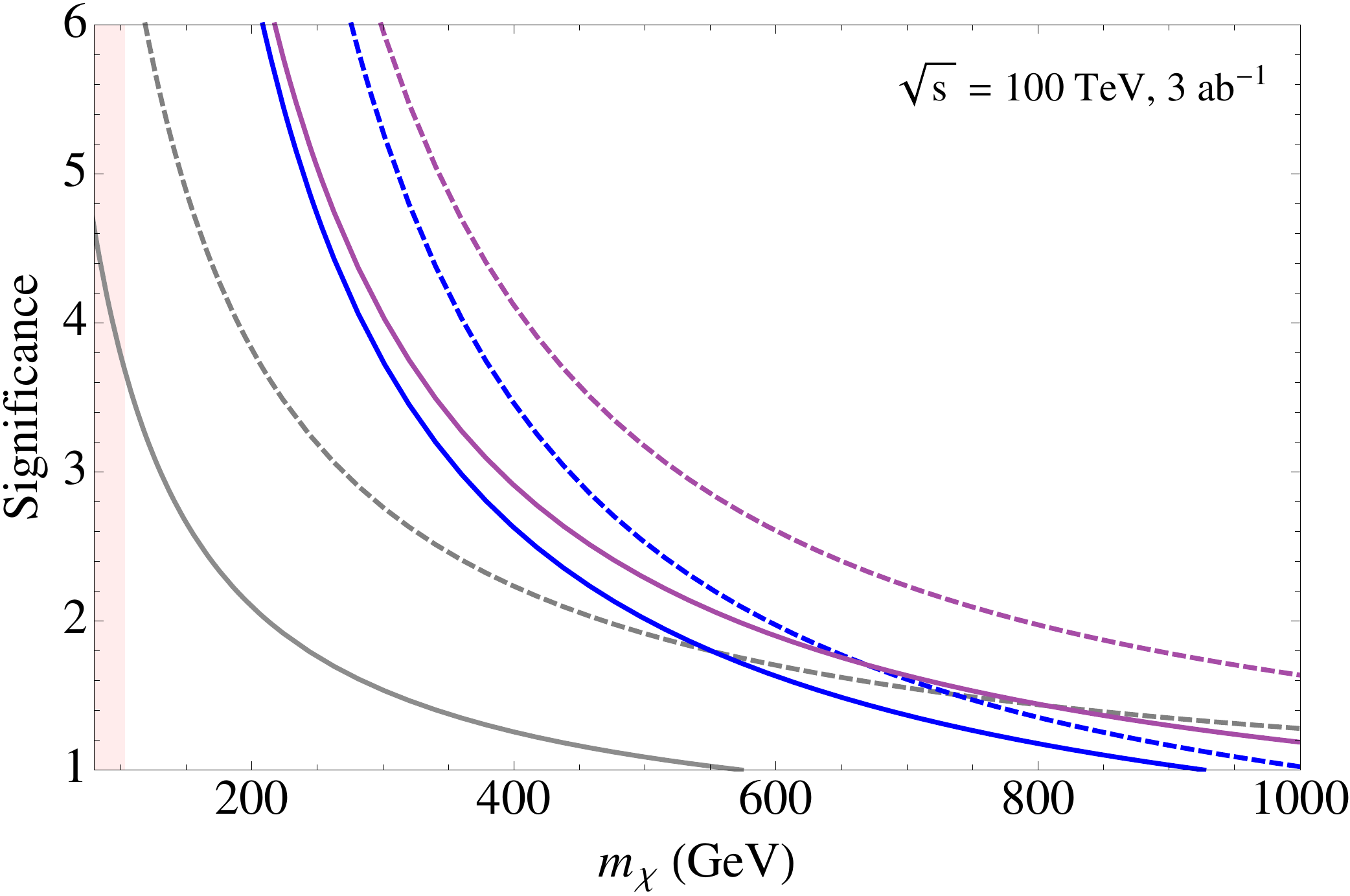}
\caption{Projected signal significance for the Higgsino doublet model with 3 ab$^{-1}$ of integrated luminosity at a future 100 TeV proton-proton collider. The curves have the same meaning as in Fig.~\ref{fig:moneyplot14tev}.}
\label{fig:moneyplot100tev}
\end{figure}

 At higher energies, the $\chi^\pm$ states are more highly boosted and emit more copious collinear radiation, improving the search prospects. In Fig.~\ref{fig:moneyplot100tev}, we illustrate the expected reach of a 100 TeV proton proton collider with 3 $\mathrm{ab}^{-1}$ of integrated luminosity. Again, we see that the $\gamma+j+\slashed{E}_{\rm T}$ search offers improved sensitivity to Higgsinos and a $2\sigma$ reach of $M_\chi\lesssim500$ GeV for 5\% systematics, nearly 300 GeV larger than the monojet search alone. The combination could be sensitive to $M_\chi\lesssim570$ GeV. If systematics can be reduced to 2\%, including $\gamma+j+\slashed{E}_{\rm T}$ could allow the Higgsino sensitivity to go up to nearly 800 GeV. Naturalness is typically stressed beyond such scales, though SUSY-breaking Higgsino mass terms can lead to lower fine-tuning values \cite{Ross:2016pml}. It may not be possible to identify photons with $E_{\rm T}>15$ GeV at such a high-energy collider; however, we find that the significance at a 100 TeV collider only degrades by approximately  $10-20\%$ when requiring $E_{\rm T}>50$ GeV, and by approximately  30\% with $E_{\rm T}>100$ GeV. Consequently, even with a higher photon $E_{\rm T}$ cut, the expected performance of the $\gamma+j+\slashed{E}_{\rm T}$ search exceeds that estimated for the $j+\slashed{E}_{\rm T}$ search.

\begin{figure}[t]
\centering
\includegraphics[width=0.45 \textwidth ]{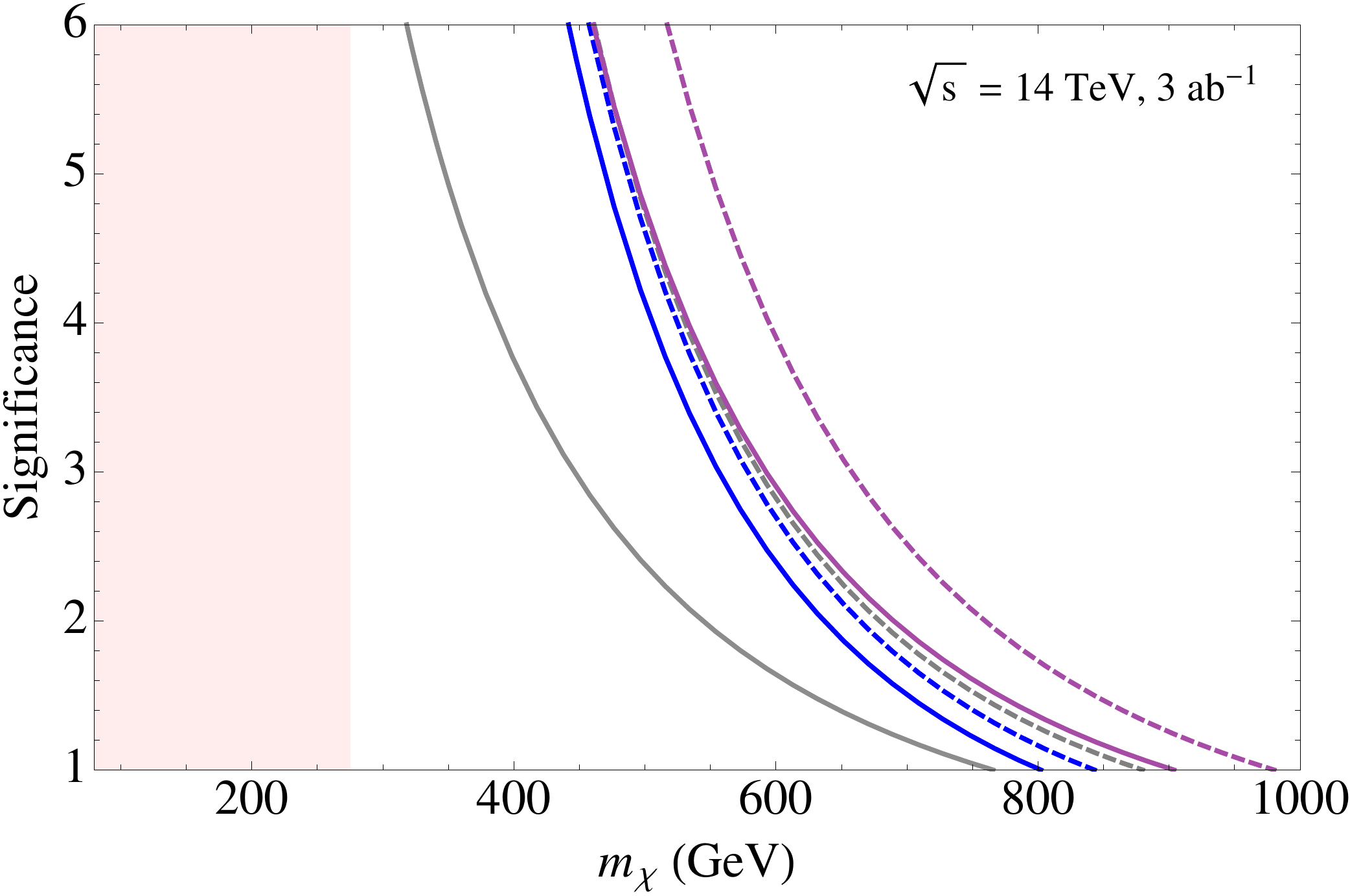}
\caption{Projected signal significance for the quintuplet model with 3 ab$^{-1}$ of integrated luminosity at the HL-LHC. The curves have the same meaning as in Fig.~\ref{fig:moneyplot14tev}. The shaded region is excluded by the 8 TeV monojet search by CMS.}
\label{fig:moneyplot14tev_5}
\end{figure}

\begin{figure}[t]
\centering
\includegraphics[width=0.45 \textwidth ]{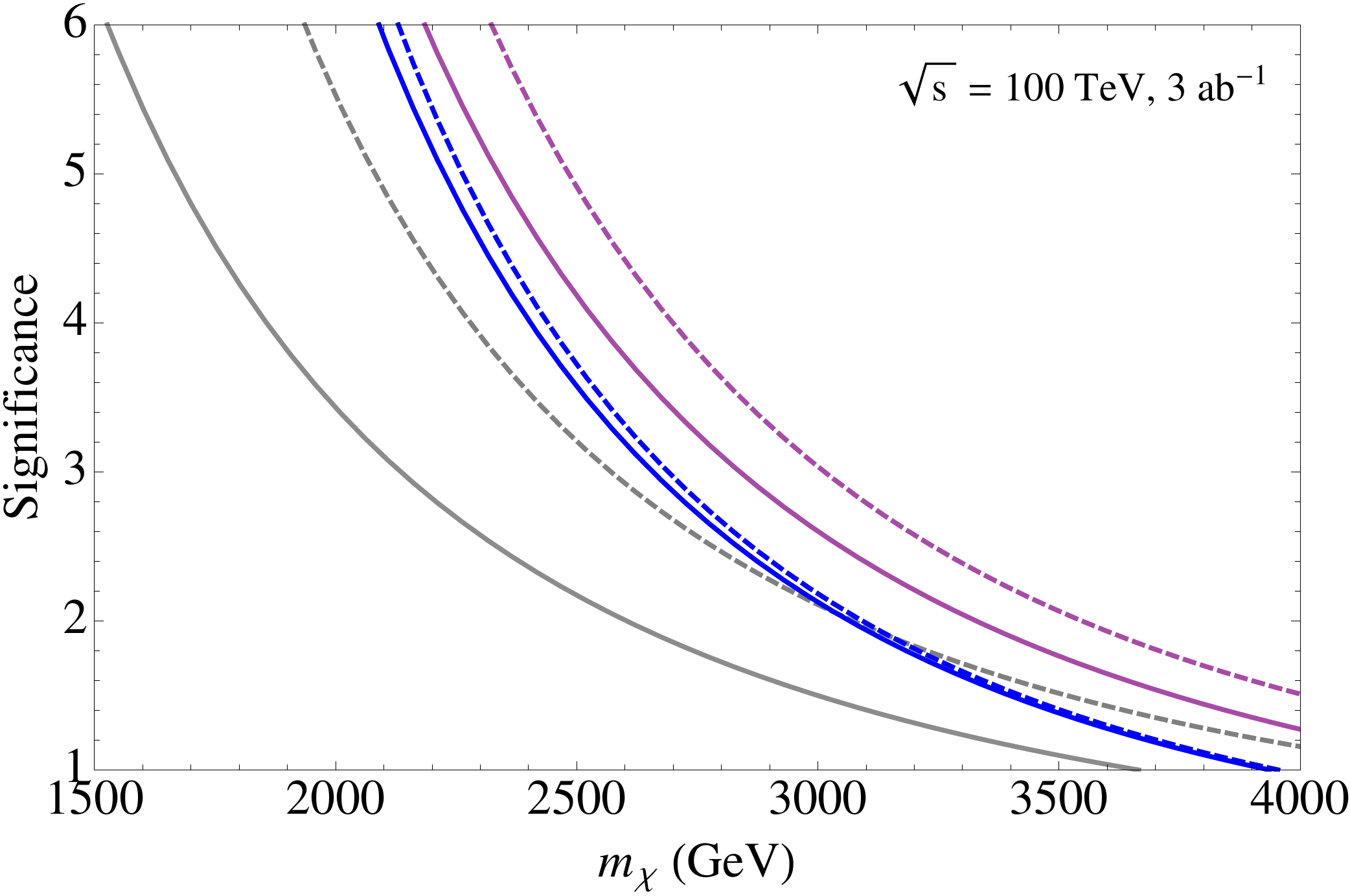}
\caption{Projected signal significance for the quintuplet model with 3 ab$^{-1}$ of integrated luminosity at a future 100 TeV $p-p$ collider. The curves have the same meaning as in Fig.~\ref{fig:moneyplot14tev}.}
\label{fig:moneyplot100tev_5}
\end{figure}

Figs.~\ref{fig:moneyplot14tev_5} and \ref{fig:moneyplot100tev_5} show the analogous expected sensitivity of the $\gamma+j+\slashed{E}_{\rm T}$ search to the quintuplet model. Since the electroweak charge is greater than the Higgsino and there are more states relative to the doublet model, both the $\gamma+j+\slashed{E}_{\rm T}$ and monojet searches give superior sensitivity to the quintuplet relative to the doublet. Indeed, because the $\chi$ production cross section is higher, existing LHC missing momentum searches already constrain the quintuplet model beyond the LEP limit. The 8 TeV monojet searches by CMS \cite{Khachatryan:2014rra} and ATLAS \cite{Aad:2015zva} set a limit of $M_\chi\lesssim 275$ GeV when recast in terms of the quintuplet. The proposed $\gamma+j+\slashed{E}_{\rm T}$ analysis has the potential to achieve an even greater significance relative to the monojet analysis due to the $Q^2$ enhancement of the doubly charged quintuplet states. Even if systematic uncertainties are not improved considerably from current values, a $\gamma+j+\slashed{E}_{\rm T}$ search can be used in conjunction with the monojet analysis to achieve sensitivity to quintuplets as heavy as 750 GeV at the HL-LHC. Moreover, the $\gamma+j+\slashed{E}_{\rm T}$ search implemented at a future 100 TeV $p-p$ collider could give a substantial improvement to the monojet search, and a combination may have sensitivity to $M_\chi\lesssim3.3-3.5$ TeV. Sensitivity to the quintuplet model is also possible with lower integrated luminosity:~$2\sigma$ sensitivity to $M_\chi\approx300$ GeV is possible with the $\gamma+j+\slashed{E}_{\rm T}$ search and $20\,\,\mathrm{fb}^{-1}$, with sensitivity to $M_\chi \approx400$ GeV in combination with a monojet + missing momentum search.

\section{Discussion and Conclusions}
\label{sec:summary}

In this article, we have proposed a new, generic method of searching for new electroweak states at hadron colliders. We demonstrated that, when a soft final-state photon is radiated from a charged electroweak state that subsequently decays into mostly invisible particles, the kinematics of this topology are sufficiently distinct from SM backgrounds to substantially improve sensitivity over existing searches.

We projected the results of this search for two representative signal models, namely a pure Higgsino doublet and an electroweak quintuplet. The former can be readily realized in common extensions of the SM, such as natural weak-scale SUSY, while the latter is a standard minimal DM candidate. However, the $\gamma+j+\slashed{E}_{\rm T}$ search is applicable to any electroweak final state that decays largely invisibly.

The new class of searches that we propose relies on the achievement of systematic uncertainties for the $j+\gamma+\slashed{E}_{\rm T}$ signature that are somewhat comparable to those for the conventional monojet searches. Fortunately, the final state we consider is also amenable to data-driven estimates of SM backgrounds similar to current methods of estimating backgrounds for $j+\slashed{E}_{\rm T}$ searches. In particular, one suitable control sample that may be used to estimate the $j \gamma (Z\rightarrow \bar\nu\nu)$ background in the signal region is obtained by replacing the hard $Z$ with a photon, obtaining a jet + high $E_{\mathrm T}$ $\gamma$ + low $E_{\mathrm T}$ $\gamma$ + $\slashed{E}_{\rm T}$ final state. Alternatively, the control region  $j \gamma (Z\rightarrow \ell \ell)$ could be defined, at the expense of poorer statistics in the control region and additional uncertainties on modeling the photon contamination from $Z\rightarrow \ell\ell$ decays that is absent in the signal region. Additionally, the $j\gamma \ell \nu$ background, in which the lepton is missed or misidentified as a jet, can be estimated by studying the control region resulting  from inverting the lepton veto and/or transverse mass cut; the correlation of the lepton kinematics with the photon $E_{\rm T}$ should be well modeled in Monte Carlo simulations.

New electroweak multiplets are generic features of physics beyond the SM. In spite of the presence of new charged states, the near degeneracy of members of the multiplet can make them extraordinarily challenging to separate from electroweak backgrounds.  As we have proposed, however, the distinctive signatures arising from the imprint of decayed charged particles, such as soft and collinear final-state photon radiation, can be leveraged to improve the experimental sensitivity to these particles. With further development of such strategies, the gaps in which new weak-scale particles can hide from collider searches continue to narrow.

\emph{Acknowledgments:} We are grateful to Valentin Hirschi, Wai-Yee Keung, Bryan Ostdiek, Stefan Prestel, and Scott Thomas for helpful conversations. We particularly thank Bryan Ostdiek for providing us with UFO files for the quintuplet model. AI thanks the Galileo Galilei Institute for Theoretical Physics and Perimeter Institute for their hospitality,
and INFN for partial support, during the completion of this work. The work of AI is supported in part by the U.S. Department of Energy under Grants DE-AC02-06CH11357 and DE-FG02-12ER41811. This research was supported in part by Perimeter Institute for Theoretical Physics. Research at Perimeter Institute is supported by the Government of Canada through Industry Canada and by the Province of Ontario through the Ministry of Economic Development \& Innovation.

\bibliographystyle{apsrevM}
\bibliography{higgsino}

\end{document}